\documentstyle[aps,prl,twocolumn,epsf]{revtex}
\begin{document}
\draft
\twocolumn[\hsize\textwidth\columnwidth\hsize\csname @twocolumnfalse\endcsname
\title{Parallel Magnetic Field Induced Transition in Transport in the Dilute
Two-Dimensional Hole system in GaAs}
\author{Jongsoo Yoon$^{1}$, C. C. Li$^{1}$, D. Shahar$^{2}$ D. C. Tsui$^{1}$, 
and M. Shayegan$^{1}$}
\address{$^{1}$Electrical Engineering Department, Princeton University, Princeton, NJ08544}
\address{${2}$Department of Condensed Matter Physics, Weizmann Institute, Rehovot 76100, Israel}
\date{\today}
\maketitle
\begin{abstract}
A magnetic field applied parallel to the two-dimensional hole system in the 
GaAs/AlGaAs
heterostructure, which is metallic in the absence of an external magnetic 
field, can drive the system into insulating at a finite field through a well 
defined transition. The value of resistivity at the transition is found to 
depend strongly on density
\end{abstract}
]

Several years ago, Kravchenko \textit{et al.} \cite{1} observed that the 
resistivity ($\rho$) of the high mobility two-dimensional (2D) electron gas in 
their Si metal-oxide-semiconductor field-effect transistor (MOSFET) samples, 
decreased by almost an order of magnitude when they lowered their sample 
temperature ($T$) below about 2K. Their observation of such a metallic 
behavior contradicts the scaling theory of localization which \cite{2} 
predicts that, in the absence of electron-electron interaction, all states in 
2D are localized in the $T\rightarrow$0 limit and that only an insulating 
phase characterized by an increasing $\rho$ with decreasing $T$ is possible at 
low $T$. They studied the $T$ dependence of $\rho$ as a function of the 2D 
carrier density ($p$) and demonstrated from the temperature coefficient 
(d$\rho$/d$T$) of $\rho$ a clear transition from the metallic to an 
insulating behavior at a ``critical'' density, $p_c$. This apparent 
metal-insulator transition (MIT) has since been reported in other low 
disorder 2D systems \cite{3}, and it appears to be a general phenomenon in 
low disorder dilute 2D systems where the Fermi energy is small and $r_s$ 
(the ratio of Coulomb interaction energy to Fermi energy) is large 
($\agt$10). To date, despite the large number of 
experimental \cite{3,4,5,6,7,8,9,10} and theoretical \cite{11,12} papers on 
this zero field MIT in the literature, there is still no consensus on the 
physics and the mechanisms behind this metallic behavior. 

Two factors that strongly influence this MIT have become apparent from the 
more recent experiments. First, the application of a magnetic field parallel 
to the 2D system ($B_{\parallel}$) induces a drastic response \cite{8,9,10}. A 
giant positive magetoresistance is observed in both the metallic and the 
insulating phases, varying continuously across the transition. In the case of 
Si MOSFET's, Simonian \textit{et al.} \cite{8} have made a detailed study 
of the temperature and electric field dependences on the magnetoresistance and
concluded that ``in the $T\rightarrow$0 limit the metallic behavior is 
suppressed by an arbitrarily weak magnetic field''. Since $B_{\parallel}$ 
couples only to the carrier's spin and does not affect its orbital motion, 
the spin degree of freedom must play a crucial role in the electronic 
processes that give rise to transport in both phases.

The second factor that has become increasingly clear is the importance of the 
role played by disorder. A close examination of all 2D systems that show the 
metallic behavior and thus the MIT as the carrier density is reduced reveals 
that $p_c$ is lower in systems with a higher mobility (i.e., a lower carrier 
scattering rate) \cite{5}. Typically, the 2D electron system (2DES) in a high 
quality Si MOSFET has a peak mobility of 
$1\times10^{4}-5\times10^{4} cm^{2}/Vs$ and 
$p_c\approx1\times10^{11} cm^{-2}$. On the other hand, the 2D hole 
system (2DHS) in GaAs/AlGaAs heterostructures, which has a comparable 
effective mass ($m^{*}$) at low densities, usually has a peak mobility of 
about 10 times higher and $p_c\approx1\times10^{10} cm^{-2}$. It is 
clear that $p_c$ decreases monotonically with decreasing disorder in the 2D 
system. In the ideal clean limit, it is well known that the Wigner crystal is 
the ground state in the low $p$ limit. The estimated critical density for 
Wigner crystallization is approximately $2\times10^9 cm^{-2}$ for the 
2DHS in GaAs, using $m^{*}=0.18m_{e}$ ($m_e$ being the free electron mass) and 
a value of 37 for $r_s$ at the crystallization \cite{13,14}.

We have recently investigated the transport properties of a 2DHS in the 
GaAs/AlGaAs heterostructure, which has an unprecedentedly high peak mobility 
of $7\times10^{5} cm^{2}/Vs$, and observed a zero field MIT at 
$p_c=7.7\times10^{9} cm^{-2}$ \cite{5}. The mobility of this 2DHS is over 
25 times that of the 2DES in the Si MOSFET whose parallel magnetic field 
response has been most extensively studied in references 8 and 9. In view of 
the fact that it is not yet clear what specific role the spins play in the 
two transport regimes and how the small amount of random disorder in high 
quality 2D systems influences the MIT, we have systematically studied the 
effect of a $B_{\parallel}$ on the transport in this high quality 2DHS. We 
find that for $p>p_{c}$, the metallic behavior persists to our lowest $T$ of 
50mK until $B_{\parallel}$ reaches a well defined ``critical'' value 
$B_{\parallel}^{c}$, beyond which the 2DHS shows an insulating behavior. At 
$B_{\parallel}^{c}$, $\rho$ is independent 
of $T$. The nonlinear I-V characteristics across this $B_{\parallel}$ induced 
transition is found to be the same as those across the zero field MIT. Below, 
we describe in more detail the changes in the transport properties of the 
2DHS under the influence of a $B_{\parallel}$ and report our observation of 
this $B_{\parallel}$ induced MIT.

We used the 2DHS created in a Si modulation doped GaAs/Al$_{x}$Ga$_{1-x}$As 
heterostructure grown on the (311)$A$ surface of an undoped GaAs substrate by 
molecular beam epitaxy. The samples were Hall bars along the 
[$\overline {2}$33] crystalographic direction, and the measurements were made 
using a dilution refrigerator in the $T$ range from 50mK to 1.1K and under 
$B_{\parallel}$ up to 14T. The hole density was tuned by a back gate in the 
range $5.7\times10^{9}<p<4.1\times10^{10} cm^{-2}$.

In Fig. 1(a), the $T$ dependence of $\rho$ in the zero field metallic phase is 
shown on a semilog plot for a hole density $p=3.7\times10^{10} cm^{-2}$ 
at several different $B_{\parallel}$'s. The bottom trace taken at 
$B_{\parallel}=0$ clearly shows a positive $d\rho/dT$, which is 
characteristic of metallic-like transport. This metallic behavior is found 
to persist to our lowest $T$ of 50mK in a magnetic field of up to about 4T. 
As $B_{\parallel}$ increases from zero, the strength of the metallic behavior 
measured by the total change in $\rho$ from about 1K to 50mK weakens 
progressively, and for $B_{\parallel}\geq4.5$T $d\rho/dT$ becomes negative. We 
take this negative $d\rho/dT$ as an indication that the 2DHS is insulating, 
and phenomenologically identify the two distinct transport regimes as a 
``metallic phase'' and an ``insulating phase''. It is clear from the figure 
that there exists a ``critical'' field $B_{\parallel}^{c}$ near 4T where 
$\rho$ becomes $T$ independent, separating the metallic and insulating 
phases. Another way of demonstrating the existence of a well defined 
$B_{\parallel}^{c}$ is to plot $\rho$ against $B_{\parallel}$ at several 
different $T$'s. Such a plot is shown in Fig. 1(b) for 
$p=1.5\times10^{10} cm^{-2}$. In this plot, the crossing point marked by the 
arrow defines $B_{\parallel}^{c}$. This is the direct consequence of the fact 
that $\rho$ decreases with decreasing $T$ in the metallic phase 
($B_{\parallel}<B_{\parallel}^{c}$), increases in the insulating phase 
($B_{\parallel}>B_{\parallel}^{c}$), and is independent of $T$ at 
$B_{\parallel}^{c}$.

Across the zero field MIT, the differential resistivity ($dV/dI$) is known to 
show an increase with increasing voltage ($V$) in the metallic phase and a 
decrease in the insulating phase \cite{4,5}. In Fig. 1(c), the $dV/dI$ 
measured at 50mK across the $B_{\parallel}$ induced MIT for 
$p=3.7\times10^{10} cm^{-2}$ is shown at similar $B_{\parallel}$'s as in Fig. 
1(a). It is clear that in the metallic phase ($B_{\parallel}<4.2$ T) $dV/dI$ 
increases as $\vline$$V$$\vline$ increases, and in the insulating phase 
($B_{\parallel}>4.2$T) it decreases. At $B_{\parallel}=4.2$T which is the 
$B_{\parallel}^{c}$, $dV/dI$ is constant, implying a linear $I-V$. This result 
also shows that there is a well defined critical field $B_{\parallel}^{c}$ 
separating the metallic and insulating phases in the presence of a parallel 
magnetic field. 

We have measured $B_{\parallel}^{c}$ and the ``critical'' resistivity 
($\rho_{c}$) as a function of $p$ in the $p>p_{c}$ regime for two samples 
cut from the same wafer, and the results are shown in Fig. 2(a)-(c). Figures 
2(a) and 2(b) show that $B_{\parallel}^{c}$ decreases with decreasing $p$ and 
approaches zero as $p$ is reduced towards $p_{c}$ where the zero field MIT 
is observed. When $B_{\parallel}^{c}$ is plotted against $p-p_{c}$ on a 
log-log scale (Fig. 2(a)), all data from the two samples form two parallel 
lines, showing that $B_{\parallel}^{c}\propto(p-p_{c})^{\alpha}$ with 
$\alpha\approx0.7$ for both samples. It is interesting to note that Hanein 
\textit{et al.} \cite {7} extracted from the $T$ dependence of $\rho$ in the 
zero field metallic phase in a previous experiment, an energy scale, $T_{0}$, 
in the form of an activation energy. Their $T_{0}$ depends linearly on $p$ in 
the range $p>2\times10^{10} cm^{-2}$ and extrapolates to zero at $p=0$. We 
postulate that the magnetic energy at $B_{\parallel}^{c}, 
g\mu_{B}B_{\parallel}^{c}$ ($g$ being the $g$-factor of holes in GaAs and 
$\mu_{B}$ the Bohr magneton), is equivalent to their $k_{B}T_{0}$ ($k_{B}$ is 
the Boltzmann constant), and compare the $p$ dependences of 
$B_{\parallel}^{c}$ and $T_{0}$ by replotting $B_{\parallel}^{c}$ against $p$ 
on a linear scale in Fig. 2(b). We find that our data from both samples in 
the range $p>2\times10^{10} cm^{-2}$ fall on straight lines that extrapolate 
to zero at $p=0$, and therefore the dependence of $B_{\parallel}^{c}$ on $p$ 
is similar to that of $T_{0}$ on $p$ in the same $p>2\times10^{10} cm^{-2}$ 
range. If we equate our $g\mu_{B}B_{\parallel}^{c}$ with their $k_{B}T_{0}$ at 
$p>2\times10^{10} cm^{-2}$, we obtain a $g$-factor of 0.1, which is of the 
same order as the hole $g$-factor of a $100\AA$ wide GaAs/AlGaAs quantum 
well \cite{15}. However, we are not able to distinguish whether the $p$ 
dependence of $B_{\parallel}^{c}$ for $p>2\times10^{10} cm^{-2}$ is indeed 
linear or of the power law form because the density range covered in our 
measurements is small.

The ``critical'' 2D resistivity $\rho _{c}$ at the transition depends strongly 
on $B_{\parallel}^{c}$ and therefore on $p$. At $p=p_{c}$, where 
$B_{\parallel}^{c}=0$, $\rho _{c}$ is of the order of one resistance quantum, 
$\rho_{Q}=h/e^{2}$ (where $h$ is Plank's constant and $e$ the electron 
charge). For $p>p_{c}$, $\rho_{c}$ decreases steeply as $B_{\parallel}^{c}$ 
increases and drops to $\sim0.03\rho_{Q}$ at $B_{\parallel}^{c}\approx7$T, 
as shown by the solid circles in Fig. 3. Fig. 2(c) shows the $\rho_{c}$ data 
from both samples as a function of $p-p_{c}$, and it is clear that for 
$p-p_{c}>2\times10^{9} cm^{-2}$ $\rho_{c}$ decreases exponentially with 
increasing $p$. This strikingly strong dependence of $\rho_{c}$ on $p$ is not 
anticipated within the MIT framework. It suggests that the observed 
insulating behavior for $B_{\parallel}>B_{\parallel}^{c}$ cannot be the 
result of thermally activated processes in a simple Anderson type of 
insulator. However, it is reminiscent of the magnetic field driven 
superconductor-insulator transition reported by Yazdani and 
Kapitulnik \cite{16} in their experiments on thin films of amorphous MoGe, 
where similar decrease in critical resistivity with increasing critical $B$ 
field is observed. They have attributed this lack of universality in their 
critical 2D resistivity to the presence of conduction by unpaired electrons. 
In this context, we should also note that Phillips \textit{et al.} \cite{11} 
have proposed that the metallic behavior observed in high mobility 2D systems 
is that of a superconductor, and the existence of a critical $B$ field is to 
be expected. However, $\rho$ in the metallic regime is known to saturate to a 
finite value instead of vanishing as $T\rightarrow0$, and the relation of the 
metallic behavior to superconductivity is not known at present.

Now, we turn to the discussion on the overall in-plane magnetoresistance. In 
Fig. 3, the in-plane magnetoresistance in the zero $B$ field insulating 
phase ($p<p_{c}$) is shown as open circles and in the metallic phase 
($p>p_{c}$) as solid lines. Regardless of whether the zero field transport is 
metallic or insulating, we observe a strong positive magnetoresistance. 
According to the $B_{\parallel}$ dependence of $\rho$, we can divide the 
entire $\rho-B_{\parallel}$ plane into two regimes: a low field regime and a 
high field regime. In the low field regime, we find that the 
magnetoresistance is well described by 
$\rho=\rho_{0}exp(B_{\parallel}^{2}/B_{0}^{2})$, where $\rho_{0}$ and 
$B_{0}$ are the fitting parameters. The value of $B_{0}$ is shown as the 
solid circles in Fig. 4 as a function of $p$. $B_{0}$ decreases as $p$ is 
reduced towards $p_{c}$ (marked by the arrow in Fig. 4) reflecting that the 
$B_{\parallel}$ dependence of $\rho$ becomes stronger. However, it is clearly 
visible in Fig. 4 that $B_{0}$ saturates to a constant value of $\sim3$T as 
the 2DHS is brought into the zero field insulating phase. The $B_{\parallel}$ 
induced transition occurs in this low field regime, and the measured 
$B_{\parallel}^{c}$'s are marked by the solid circles in Fig. 3 (the dashed 
line is a guide to the eye). As the magnetic field is increased beyond 
$B_{\parallel}^{*}$ (which is indicated by the dotted line in Fig. 3) into he 
high field regime, the dependence of $\rho$ on $B_{\parallel}$ changes. In 
this regime, the magnetoresistance is of the form 
$\rho =\rho _{1}exp(B_{\parallel}/B_{1})$, where 
$\rho _{1}$ and $B_{1}$ are the fitting parameters. As shown by the open 
circles in Fig. 4, the $p$ dependence of $B_{1}$ is similar to that of 
$B_{0}$; $B_{1}$ also decreases as $p$ is reduced towards $p_{c}$ and 
saturates to $\sim 1.5$T at $p<p_{c}$. While the overall magnetoresistance 
evolves smoothly as $p$ is changed across $p_{c}$, it is obvious from Fig. 3 
that $B_{\parallel}^{*}$ (the boundary separating the low field and high field 
regimes) decreases with decreasing $p$ in the zero field metallic phase, but 
is independent of $p$ in the insulating phase. It is interesting to note that 
all three characteristic fields for the in-plane magnetoresistance, 
$B_{0}$, $B_{1}$, and $B_{\parallel}^{*}$, become independent of $p$ when $p$ 
is reduced below $p_{c}$.

The low field magnetoresistance at $B_{\parallel}<B_{\parallel}^{*}$ is very 
similar to that observed in the Si MOSFET's \cite{8,9,10}, where a positive 
magnetoresistance at low fields is followed by a saturation at high fields. 
Mertes \textit{et al.} \cite{9} have interpreted the low field positive 
magnetoresistance in Si MOSFET's by the hopping model of Kurobe and Kamimura 
\cite{17}, in which hopping becomes more difficult as more spins get aligned 
with $B_{\parallel}$. This interpretation is supported by the observation of 
Okamoto \textit{et al.} \cite{10} that the field where magnetoresistance 
starts to saturate coincides with the field expected for complete spin 
alignment. Such a hopping model can also be applied to explain our data in 
the $p<p_{c}$ insulating regime. However, for $p>p_{c}$, transport is 
metallic and hopping is not relevant; some other mechanisms involving spins 
must be operative. 

The exponential divergence of the magnetoresistance observed at high fields 
($B_{\parallel}>B_{\parallel}^{*}$), on the other hand, has not been observed 
in other 2D systems before, and cannot be explained by existing theoretical 
models. The model by Lee and Ramakrishnan \cite{18} for a weakly disordered 
system, of which the in-plane magnetoresistance arises from spin splitting, 
predicts a logarithmic divergence. In the hopping model by Kurobe and 
Kamimura, the magnetoresistance is expected to saturate when all spins are 
aligned. In our data, the exponential dependence of $\rho$ on $B_{\parallel}$ 
is observed up to our highest field of 14T, and $B_{0}$, as seen in Fig. 4, 
varies continuously with $p$ across $p_{c}$. Also, it appears that the 
influence of the parallel magnetic field on the energy structure of our 2DHS 
is not the cause of such strong but simple $B_{\parallel}$ dependence in both 
transport regimes. It is possible that this exponential divergence of the 
magnetoresistance is a phonomenon characteristic of new electronic processes 
in the 2D system in its clean limit.

We thank R. Bhatt, P. Phillips, M. Hilke, S. Papadakis, and Y. Hanein for 
fruitful discussions. This work is supported by the NSF.





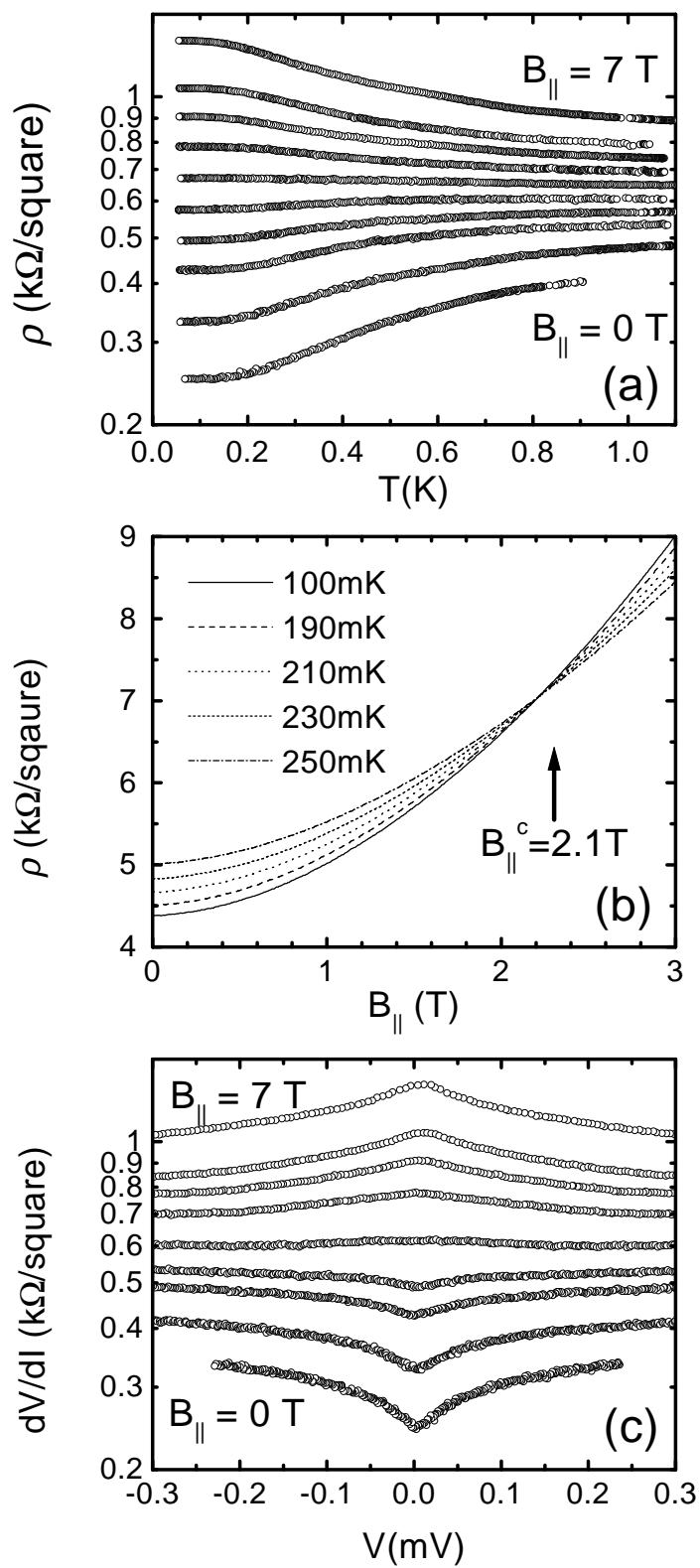
Fig. 1. (a) $T$ dependence of $\rho$ at $p=3.7\times10^{10} cm^{-2}$ and at 
$B_{\parallel}$=0, 2, 3, 3.5, 4, 4.5, 5, 5.5, 6, and 7T from the bottom. 
(b) $\rho$ is plotted as a function of $B_{\parallel}$ 
at $p=1.5\times10^{10} cm^{-2}$ and at five different temperatures. The 
crossing point defines the $B_{\parallel}^{c}$ at 2.1T as marked by the 
arrow. (c) Differential resistivity ($dV/dI$) is plotted against $V$ at 50mK 
and $p=3.7\times10^{10} cm^{-2}$, at magnetic fields, $B_{\parallel}$=0, 
2, 3, 3.5, 4.2, 5, 5.5, 6, and 7T from the bottom.

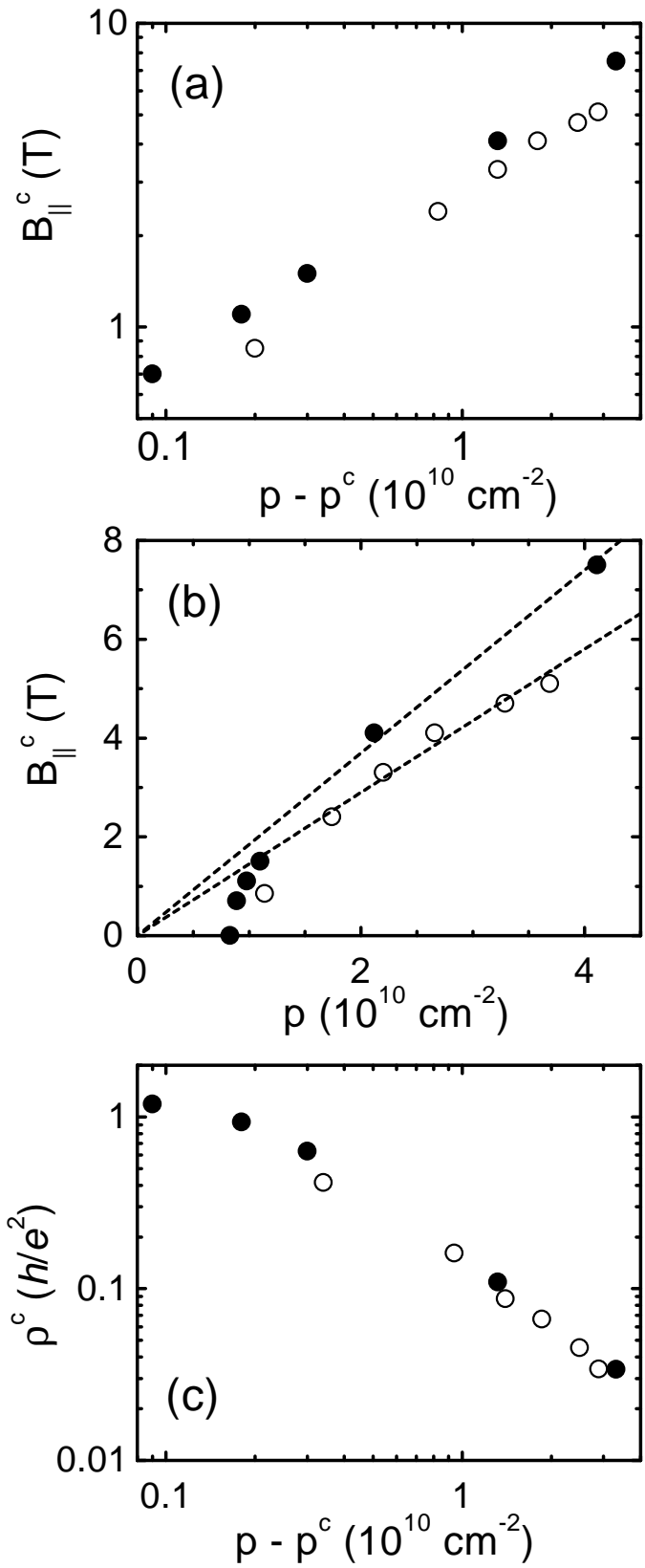
Fig. 2. (a) $B_{\parallel}^{c}$ is plotted against $p-p^{c}$. The solid and 
open circles are for two samples cut from the same wafer. 
(b) $B_{\parallel}^{c}$ vs $p$ in a linear scale. The dotted lines are to 
indicate that $B_{\parallel}^{c}$ is approximately linear in the 
range $p>2\times10^{10} cm^{-2}$ extrapolating to zero at $p=0$. 
(c) $\rho^{c}$ is shown as a function of $p-p^{c}$.

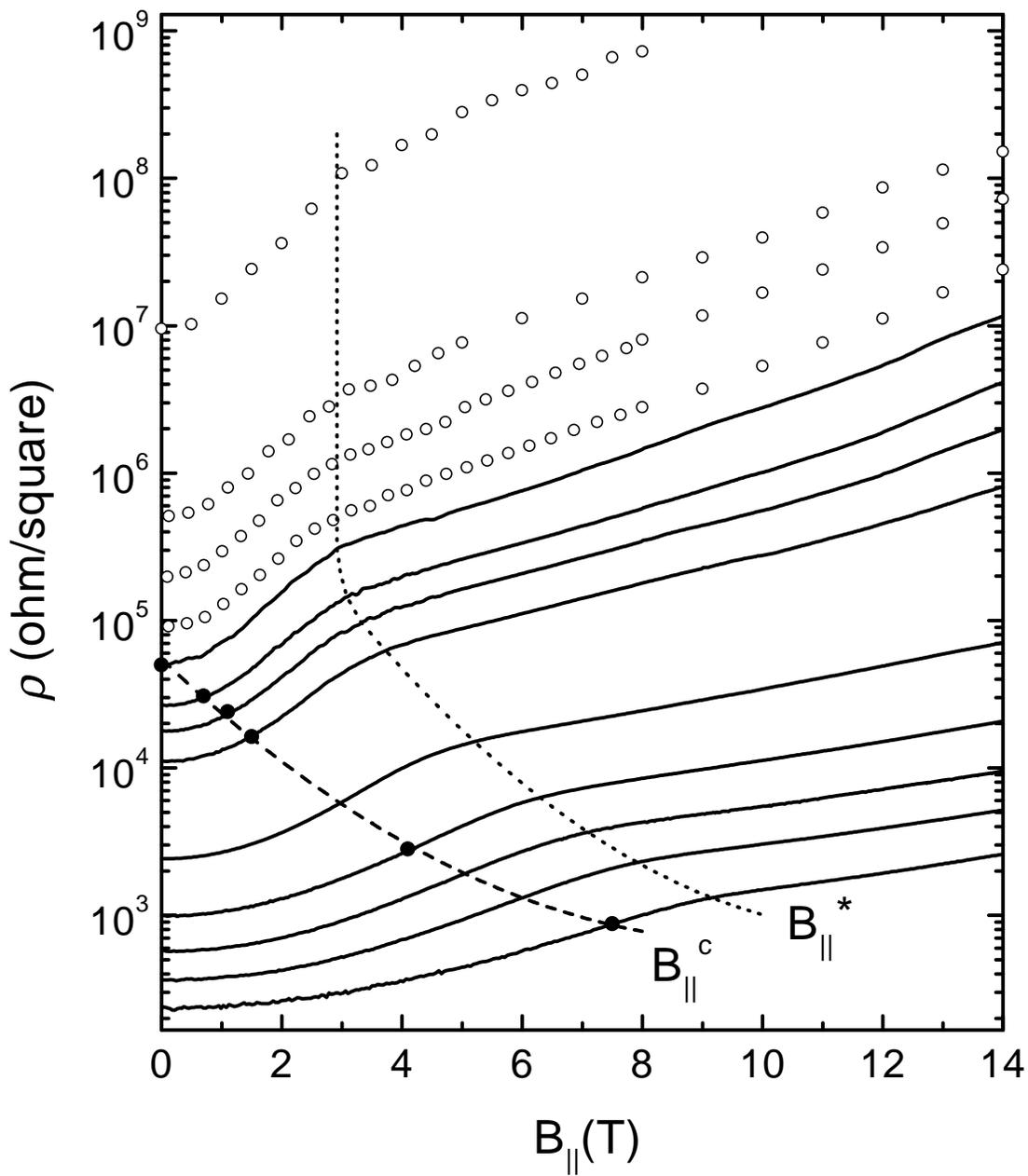
Fig. 3. The $B_{\parallel}$ dependence of $\rho$ is shown at 50mK at hole 
densities, from the bottom. 4.11, 3.23, 2.67, 2.12, 1.63, 1.10, 0.98, 0.89, 
0.83, 0.79, 0.75, 0.67, and 0.57$\times10^{10} cm^{-2}$. The solid lines are
for $p>p^{c}$ and the open circles for $p<p^{c}$. The colid circles denote 
experimentally determined $B_{\parallel}^{c}$'s, and the dashed line is a 
guide to the eye. $B_{\parallel}^{*}$, the boundary separating the low field 
and the high field regimes, is marked as the dotted line.

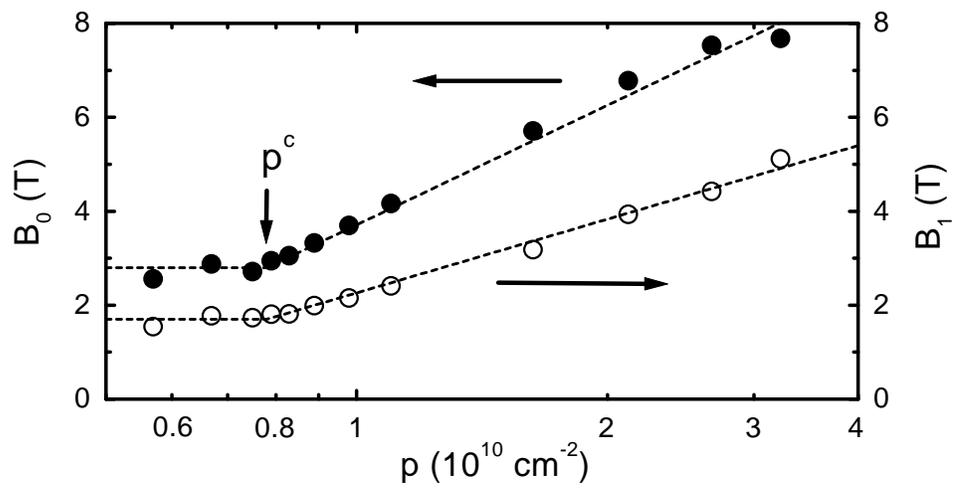
Fig. 4. $B_{0}$ and $B_{1}$, obtained from fitting the data in Fig. 3 in the 
form $\rho=\rho_{0}exp(B_{\parallel}^{2}/B_{0}^{2})$ for the low field regime 
and $\rho=\rho_{1}exp(B_{\parallel}/B_{1})$ for the high field regime, are 
plotted as a function of $p$. The dashed lines are guides to the eye. 
The ``critical'' density $p^{c}$ is marked by the arrow


\begin{references}
\bibitem{1} S. V. Kravchenko, G. V. Kravchenko, J. E. Furneaux, V. M. Pudalov,
and M. K'Iorio, Phys. Rev. B {\bf 50}, 8039 (1994)
\bibitem{2} E. Abrahams, P.W. Anderson, D. C. Licciardello, and T. V. 
Ramakrishnan, Phys. Rev. Lett. {\bf 42}, 673 (1979)
\bibitem{3} D. Popovic, A. B. Fowler, and S. Washburn, Phys. Rev. Lett. 
{\bf 79} 1543 (1997); P. T. Coleridge, R. L. Williams, Y. Feng, and P. 
Zawadzki, Phys. Rev. B {\bf 56}, R12764 (1997); M. Y. Simmons, A. R. 
Hamilton, M. Pepper, E. H. Linfield, R. D. Rose, D. A.. Ritchie, A. K. 
Savchenko, and T. G. Griffiths, Phys. Rev. Lett. {\bf 80}, 1292 (1998); S. J. 
Papadakis and M. Shayegan, Phys. Rev. B {\bf 57}, R15068 (1998); S. J. 
Papadakis, E. P. De Poortere, H. C. Manoharan, M. Shayegan, and R. 
Winkler, \textit{Science} {\bf 283}, 2056 (1999)
\bibitem{4} S. V. Kravchenko, D. Simonian, M. P. Sarachik, W. Mason, and 
J. E. Furneaux, Phys. Rev. Lett. {\bf 77}, 4938 (1996); D. Simonian, S. V. 
Kravchenko, and M. P. Sarachik, Phys. Rev. B {\bf 55}, R13421 (1997)
\bibitem{5} J. Yoon, C. C. Li, D. Shahar, D. C. Tsui, and M. Shayegan, 
Phys. Rev. Lett. {\bf 82}, 1744 (1999)
\bibitem{6} A. P. Mills, A. P. Ramirez, L. N. Pfeiffer, and K. W. West, 
preprint cond-mat/9905176
\bibitem{7} Y. Hanein, U. Meirav, D. Shahar, C. C. Li, D. C. Tsui, and 
H. Shtrikman, Phys. Rev. Lett. {\bf 80}, 1288 (1998); Y. Hanein, D. Shahar, 
J. Yoon, C. C. Li, D. C. Tsui, and H. Shtrikman, Phys. Rev. B {\bf 58}, 
R13338 (1998)
\bibitem{8} D. Simonian, S. V. Kravchenko, M. P. Sarachik, and V. M. 
Pudalov, Phys. Rev. Lett. {\bf 79}, 2304 (1997)
\bibitem{9} K. M. Mertes, D. Simonian, M. P. Sarachik, S. V. Kravchenko, 
and T. M. Klapwijk, preprint cond-mat/9903179
\bibitem{10} T. Okamoto, K. Hosoya, S. Kawaji, and A. Yagi, Phys. Rev. 
Lett. {\bf 82}, 3875 (1999)
\bibitem{11} P. Phillips, Y. Wan, I. Martin, S. Knysh, and D. 
Dalidovich, \textit{Nature}, {\bf 395}, 253 (1998)
\bibitem{12} V. M. Pudalov, JETP Lett. {\bf 66}, 175 (1997); V. 
Dobrosavljevic, E. Abrahams, E. Miranda, and S. Chakravarty, Phys. Rev. 
Lett. {\bf 79}, 455 (1997); S. He and X.C. Xie, Phys. Rev. Lett. {\bf 80}, 
3324 (1998); Q. Si and C. M. Varma, Phys. Rev. Lett. {\bf 81}, 4951 (1998); 
C. Castellani, C. Di Castro, and P. A. Lee, Phys. Rev. B {\bf 57}, R9381 
(1998); S. Chakravarty, L. Y. Lin, and E. Abrahams, Phys. Rev. B {\bf 58}, 
R559 (1998); S. Chakravarty, S. Kivelson, C. Nayak, and K. Voelker, 
preprint cond-mat/9805383; S. Das Sarma and E. H. Hwang, preprint 
cond-mat/9812216; A. Altshuler and D. L. Maslov, Phys. Rev. Lett. {\bf 82}, 
145 (1999)
\bibitem{13} B. E. Cole, J. M. Chamberlain, M. Henini, T. Cheng, W. Batty, 
A. Wittlin, J. A. A. J. Perenboom, A. Ardavan, A. Polisski, and J. 
Singleton, Phys. Rev. B {\bf 55}, 2503 (1997)
\bibitem{14} B. Tanatar and D. M. Ceperley, Phys. Rev. B {\bf 39}, 5005 (1989)
\bibitem{15} M. J. Snelling, E. Blackwood, C. J. Donagh, R. T. Harley, and 
C. T. B. Foxon, Phys. Rev. B {\bf 45}, 3922 (1992)
\bibitem{16} A. Yazdani and A. Kapitulnik, Phys. Rev. Lett. {\bf 74}, 
3037 (1995)
\bibitem{17} A. Kurobe and H. Kamimura, J. Phys. Soc. Japan, {\bf 51}, 1904 
(1982)
\bibitem{18} P. A. Lee and T. V. Ramakrishnan, Phys. Rev. B {\bf 26}, 4009 
(1982)

\end{references}
\end{document}